# Observation of stationary spontaneous Hawking radiation and the time evolution of an analogue black hole

Victor I. Kolobov, Katrine Golubkov, Juan Ramón Muñoz de Nova, and Jeff Steinhauer[1]

*Department of Physics, Technion—Israel Institute of Technology, Technion City, Haifa 3200003, Israel*

**The emission of Hawking radiation from a black hole was predicted to be stationary, which is necessary for the correspondence between Hawking radiation and black-body radiation. Spontaneous Hawking radiation was observed in analogue black holes in atomic Bose-Einstein condensates, although the stationarity was not probed. Here, we confirm that the spontaneous Hawking radiation is stationary, by observing such a system at six different times. Furthermore, we follow the time evolution of Hawking radiation and compare and contrast it with predictions for real black holes. We observe the ramp up of Hawking radiation followed by stationary spontaneous emission, similar to a real black hole. The end of the spontaneous Hawking radiation is marked by the formation of an inner horizon, which is seen to cause stimulated Hawking radiation as predicted. We find that the stimulated Hawking and partner particles are directly observable, and that the stimulated emission evolves from multi-mode to monochromatic. Numerical simulations suggest that Bogoliubov-Cherenkov-Landau stimulation predominates, rather than black-hole lasing.**

As noted by Bekenstein, there are a number of similarities between black-hole physics and thermodynamics[1]. Hawking presented one such similarity, that the radiation from a black hole should be stationary like black-body radiation[2,3]. The search for Hawking radiation from real, microscopic black holes is ongoing in particle accelerators[4,5]. Primordial microscopic black holes are also possibilities[6]. Classical, stimulated Hawking radiation was observed in water waves[7-9] as well as in a fiber-optical system[10]. Spontaneous Hawking radiation in analogue black holes was suggested[11], developed theoretically[12-23], and observed[24,25]. However, the stationarity of the Hawking radiation was not probed, since it would require observation over time. Here, we measure the spontaneous Hawking radiation at various times after the formation of the horizon, and confirm its stationarity.

[1] Jeffs@physics.technion.ac.il

An analogue black hole has regions of subsonic ($v < c$) and supersonic ($v > c$) flow separated by the horizon ($v = c$), where $v$ is the flow velocity and $c$ is the speed of sound, as illustrated in Fig. 1a. The velocity gradient is induced by a step potential at the horizon. Fig. 1b, c shows the dispersion relation outside and inside the analogue black hole. The Hawking and partner modes are indicated by circles. According to the group velocity $\partial\omega/\partial k$, the Hawking modes escape, but the partner modes are trapped inside the analogue black hole.

Spontaneous Hawking radiation occurs in the absence of incoming particles impinging on the horizon. According to Hawking's calculation, the horizon of a black hole should spontaneously emit pairs of particles, as illustrated in Fig. 1d. The Hawking particle HR exits the black hole with frequency $\omega$ and energy $\hbar\omega$. The partner particle P falls into the black hole with the same frequency but negative energy. The population of the modes is given by the thermal distribution $1/\left(e^{\hbar\omega/k_{\mathrm{B}}T_{\mathrm{H}}} - 1\right)$, where the Hawking temperature is given by the surface gravity. In the analogue case,

$$k_{\mathrm{B}}T_{\mathrm{H}} = \frac{\hbar}{2\pi}\left(\frac{dv}{dx} + \frac{dc}{dx}\right)\Bigg|_{\mathrm{horizon}} \tag{1}$$

We see that $v(x)$ and $c(x)$ play the role of the metric, and their derivatives at the horizon are the analogue of the surface gravity. Unless otherwise specified, $x$ is given in units of the healing length $\xi = \hbar/m\sqrt{c_{\mathrm{out}}c_{\mathrm{in}}}$ where $m$ is the atomic mass, and $c_{\mathrm{out}}$ and $c_{\mathrm{in}}$ are the speeds of sound outside and inside the analogue black hole, respectively. The length scale of the horizon region sets the wavelength of the Hawking radiation, and thus determines the predicted Hawking temperature. For an analogue black hole, this length scale is implicit in Eq. 1, whereas for a real non-extremal black hole, this length scale is set by the diameter of the black hole[6].

The experimental technique is described in Ref. 25. That work studied the Hawking radiation in an analogue black hole in a Bose-Einstein condensate, at approximately 124 ms after the formation of the horizon (the current work includes this data point). Here, we repeat the observation at various times after the formation of the horizon. In order to observe the Hawking radiation at each time, we gather an ensemble of density profiles and compute the density-density correlation function $G^{(2)}(x, x')$, as shown in Fig. 2. These correlation functions are computed from a total of 97,000 repetitions of the experiment, corresponding to 124 days of continuous measurement. The Hawking radiation is seen to exhibit a ramp-up period, followed by a period of spontaneous emission, and then two stimulated periods, characterized by multi-mode and monochromatic emission.

**The period of stationary spontaneous Hawking radiation**

In the period of spontaneous Hawking radiation, each of the 6 correlation functions shows a band of correlations extending from the center of the plot into the upper left and lower right quadrants. These are the correlations between the Hawking and partner quasiparticles. The negative pair correlations imply positive mutual information between the outside and inside of the analogue black hole, as can be understood from Ref. 26. In contrast, in the ramp-up period, almost no correlations of this type are visible.

The blue circles of Fig. 3a indicate the predicted Hawking temperature by Eq. 1. It is seen to be constant for all but the latest times. The spectrum of correlations is obtained from Fig. 2 by our Fourier transform technique[19,24,25]. The area under the spectrum is a measure of the quantity of Hawking radiation, as indicated by green circles in Fig. 3a. At early times, the quantity is well below the prediction (blue circles). It takes some time for the Hawking radiation to ramp up to the quantity expected for a thermal spectrum. The ramp up is similar to the dashed line, which indicates the collapse of a spherical cloud of dust with a mass of 170 solar masses[27], although we are not aware of any theoretical prediction that this should be the case. The time origin of the dashed line is a fit parameter, and we have assumed the Stefan-Boltzmann law, that the radiated flux is proportional to $T_{\mathrm{H}}^{4}$ (Ref. 6).

After the ramp-up period, the quantity of Hawking radiation is as expected for spontaneous Hawking radiation, as seen through the agreement of the green and blue circles in Fig. 3a. This agreement continues for some time, demonstrating the stationarity of the Hawking emission. The magenta pluses indicate the measured width of the spectrum of Hawking radiation. In other words, they indicate the distribution of wavenumbers or energies. During the spontaneous period, the magenta pluses are seen to be in approximate agreement with the predicted Hawking temperature indicated by the blue circles. This shows that the length scale of the horizon sets the wavelength of the emitted Hawking radiation, as expected. In summary, all three symbols agree during the spontaneous period, indicating a stationary thermal spectrum at the predicted Hawking temperature.

Fig. 4a shows the average of the 6 correlation functions of the spontaneous period. For comparison, Fig. 4b shows a theoretical model employing an idealized stationary configuration[28]. Fig. 4b has extra fringes F due to the dispersion, but no such fringes are seen in Fig. 4a. The experiment thus displays little dispersion, and the associated corrections to Fig. 3a are small. Indeed, the corrections would be on the order of the error bars in Fig. 3a, and hence they would not affect the results of this work.

**The stimulated periods**

The end of the spontaneous period is marked by the formation of the inner horizon, due to the gradual slowing of the condensate as it travels uphill after flowing over the step potential shown in Fig. 1a. The inner horizon, combined with the superluminal dispersion relation, allows for Bogoliubov-Cherenkov-Landau (BCL) stimulated Hawking radiation[29], in which stimulating waves are generated at the inner horizon[30], since it moves at the Landau critical velocity[31]. These waves have zero frequency in the reference frame of the inner horizon, as indicated by $k_0$ in Fig. 1c. In an analogue black hole such as ours, the inner horizon recedes from the outer horizon with a velocity $v_{\mathrm{IH}}$ (Ref. 29). Due to the resulting Doppler shift, $k_0$ has a finite frequency $k_0 v_{\mathrm{IH}}$ in the reference frame of the outer horizon, as indicated by the square in Fig. 1c. This wave travels to the outer horizon and stimulates Hawking radiation. The process is deterministic so naively, it would not appear in the correlation function. However, it has been suggested that noise in the form of experimental variations between runs can affect the correlation function[32]. Such variations could change the phase and/or amplitude of the BCL-stimulated Hawking radiation from run to run, allowing it to appear in the correlation function.

Just after the incoming BCL wave first impinges on the outer horizon, the resulting stimulated Hawking radiation has a multi-mode character. During this transient period of multi-mode stimulated Hawking radiation, the quantity of Hawking radiation increases greatly relative to the spontaneous level; the green circles are well above the blue circles in Fig. 3a. However, the multi-mode stimulated period is similar to the spontaneous period in the sense that the correlation pattern is non-sinusoidal, as seen in Figs. 2 and 3b. This suggests that several of the modes of the Hawking radiation are stimulated. We confirm the multi-mode nature of the stimulated Hawking radiation by contrasting Fig. 2 with single modes generated in a preliminary experiment, where a sinusoidal pattern is seen at all times, including the multi-mode stimulated period (see Methods). Although the incoming BCL wave is monochromatic for long times, it stimulates the multiple modes since very few cycles are involved. The frequency $k_0 v_{\mathrm{IH}}/2\pi$ of the incoming BCL wave is 35 Hz, using the measured $k_0$ and $v_{\mathrm{IH}}$ (see Extended Data). This wave is launched when the inner horizon forms, and requires 42 ms to propagate from the inner horizon to the outer horizon. For the three observations in the multi-mode stimulated period, the incoming BCL wave has oscillated by 0.5, 1.5, and 2.0 periods respectively, since the moment when it first impinged on the outer horizon. These values are roughly similar to the number of oscillations seen in the multi-mode profiles of Fig. 3b. The narrowing of the spectrum begins in the multi-mode period and continues in the monochromatic period, as seen in Fig. 3a since the magenta pluses drop well below the blue circles.

In the period of monochromatic stimulated Hawking radiation, the Hawking/partner correlations obtain a single-frequency sinusoidal character, as seen in Figs. 2 and 3b. The correlations also become very strong. The single modes are also seen in the Fourier transforms of Fig. 4c, d. We can check whether these modes are consistent with the BCL stimulation mechanism. The frequencies $k_0 v_{\mathrm{IH}}$ for the times of Fig. 4c, d are converted to wavenumbers outside and inside the analogue black hole via the measured dispersion relations (see Methods), as indicated by blue circles. These predicted BCL-stimulated modes are seen to be in rough agreement with the observed modes of stimulated Hawking radiation. The end of the monochromatic period marks the end of the evolution, since the outside of the analogue black hole has become rather small (see Methods).

## Ruling out black-hole lasing

In BCL stimulation, the incoming wave is deterministically produced at the inner horizon. In contrast, in black-hole lasing, the partner particles of the Hawking radiation reflect from the inner horizon, and become the "in" mode of Fig. 1c,d[33]. Black-hole lasing inherits the stochastic nature of the partner particles. Thus, we can differentiate between BCL stimulation and black-hole lasing by determining whether the stimulation is deterministic or stochastic. If the stimulation is deterministic, it should only appear in the correlation function in the presence of experimental variations between runs. Fig. 5 shows numerical simulations similar to the experiment, of the type presented in Ref. 25 except for the addition of experimental variations. Specifically, an overall slope is added to the external potential, where the slope varies from run-to-run. The third row combines both the quantum fluctuations and the experimental variations, giving a result similar to the experimental Fig. 2 for both the multi-mode and the monochromatic stimulated periods. Thus, experimental variations between runs are required to numerically reproduce the stimulated periods, which shows that the stimulation is deterministic, as expected for BCL stimulation but not black-hole lasing. In contrast, in the spontaneous period, there is no deterministic component to the Hawking radiation, so the run-to-run variations have no effect. It is not surprising that BCL stimulation dominates both the multi-mode and monochromatic periods; the continuous narrowing of the correlation spectrum seen throughout these periods (Fig. 3a, magenta pluses) suggests a single underlying process.

In conclusion, the analogue black hole exhibits four time periods with qualitatively different behaviors. During the first two periods, the Hawking radiation ramps up to a stationary, spontaneous emission with negligible effect of dispersion, in analogy with the expected evolution of a real black hole. After the spontaneous period, an inner horizon forms, which dramatically alters the further evolution. Specifically, the quantity of Hawking radiation greatly increases due to stimulation during the multi-mode and monochromatic periods. We find that the BCL-stimulated Hawking/partner pairs appear in the correlation function, an observation which was not predicted in the literature. For a real black hole, such stimulation would rely on Lorentz violation, which might occur at an energy much higher than the Planck scale[34]. The life of the analogue black hole ends with the shrinking of the region outside the horizon, rather than inside.

1. Bekenstein, J. D. Black holes and entropy. *Phys. Rev. D* **7**, 2333-2346 (1973).
2. Hawking, S. W. Black hole explosions? *Nature* **248**, 30-31 (1974).
3. Hawking, S. W. Particle creation by black holes. *Commun. Math. Phys.* **43**, 199-220 (1975).
4. Dimopoulos, S. & Landsberg, G. Black holes at the large hadron collider. *Phys. Rev. Lett.* **87**, 161602 (2001).
5. Giddings, S. B. & Thomas, S. High energy colliders as black hole factories: The end of short distance physics. *Phys. Rev. D* **65**, 056010 (2002).
6. Page, D. N. Particle emission rates from a black hole: Massless particles from an uncharged, nonrotating hole. *Phys. Rev. D* **13**, 198-206 (1976).
7. Weinfurtner, S., Tedford, E. W., Penrice, M. C. J., Unruh, W. G. & Lawrence, G. A. Measurement of stimulated Hawking emission in an analogue system. *Phys. Rev. Lett.* **106**, 021302 (2011).
8. Rousseaux, G., Mathis, C., Maïssa, P., Philbin, T. G. & Leonhardt, U. Observation of negative-frequency waves in a water tank: a classical analogue to the Hawking effect? *New J. Phys.* **10**, 053015 (2008).
9. Euvé, L.-P., Michel, F., Parentani, R., Philbin, T. G. & Rousseaux, G. Observation of noise correlated by the Hawking effect in a water tank. *Phys. Rev. Lett.* **117**, 121301 (2016).
10. Drori, J., Rosenberg, Y., Bermudez, D., Silberberg, Y. & Leonhardt, U. Observation of stimulated Hawking radiation in an optical analogue. *Phys. Rev. Lett.* **122**, 010404 (2019).
11. Unruh, W. G. Experimental black-hole evaporation? *Phys. Rev. Lett.* **46**, 1351-1353 (1981).
12. Garay, L. J., Anglin, J. R., Cirac, J. I. & Zoller, P. Sonic analog of gravitational black holes in Bose-Einstein condensates. *Phys. Rev. Lett.* **85**, 4643-4647 (2000).
13. Visser, M. Acoustic black holes: horizons, ergospheres and Hawking radiation. *Class. Quantum Grav.* **15**, 1767-1791 (1998).
14. Balbinot, R., Fabbri, A., Fagnocchi, S., Recati, A. & Carusotto, I. Nonlocal density correlations as a signature of Hawking radiation from acoustic black holes. *Phys. Rev. A* **78**, 021603(R) (2008).
15. Carusotto, I., Fagnocchi, S., Recati, A., Balbinot, R. & Fabbri, A. Numerical observation of Hawking radiation from acoustic black holes in atomic Bose-Einstein condensates. *New J. Phys.* **10**, 103001 (2008).
16. Macher, J. & Parentani, R. Black-hole radiation in Bose-Einstein condensates. *Phys. Rev. A* **80**, 043601 (2009).
17. Larré, P.-É., Recati, A., Carusotto, I. & Pavloff, N. Quantum fluctuations around black hole horizons in Bose-Einstein condensates. *Phys. Rev. A* **85**, 013621 (2012).
18. Recati, A., Pavloff, N. & Carusotto, I. Bogoliubov theory of acoustic Hawking radiation in Bose-Einstein condensates. *Phys. Rev. A* **80**, 043603 (2009).
19. Steinhauer, J. Measuring the entanglement of analogue Hawking radiation by the density-density correlation function. *Phys. Rev. D* **92**, 024043 (2015).
20. Jacobson, T. A. & Volovik, G. E. Event horizons and ergoregions in $^3$He. *Phys. Rev. D* **58**, 064021 (1998).

21. Schützhold, R. & Unruh, W. G. Hawking radiation in an electromagnetic waveguide? *Phys. Rev. Lett.* **95**, 031301 (2005).
22. de Nova, J. R. M., Guéry-Odelin, D., Sols, F. & Zapata, I. Birth of a quasi-stationary black hole in an outcoupled Bose-Einstein condensate. *New J. Phys.* **16**, 123033 (2014).
23. Balbinot, R., Fagnocchi, S., Fabbri, A. & Procopio, G. P. Backreaction in acoustic black holes. *Phys. Rev. Lett.* **94**, 161302 (2005).
24. Steinhauer, J. Observation of quantum Hawking radiation and its entanglement in an analogue black hole. *Nature Phys.* **12**, 959-965 (2016).
25. de Nova, J. R. M., Golubkov, K., Kolobov, V. I. & Steinhauer, J. Observation of thermal Hawking radiation and its temperature in an analogue black hole, *Nature* **569**, 688-691 (2019).
26. Giovanazzi, S. Entanglement entropy and mutual information production rates in acoustic black holes. *Phys. Rev. Lett.* **106**, 011302 (2011).
27. Brout, R., Massar, S., Parentani, R. & Spindel, Ph. A primer for black hole quantum physics. *Phys. Rept.* **260**, 329-446 (1995).
28. Isoard M. & Pavloff, N. Departing from thermality of analogue Hawking radiation in a Bose-Einstein condensate. *Phys. Rev. Lett.* **124**, 060401 (2020).
29. Wang, Y.-H., Jacobson, T., Edwards, M. & Clark, C. W. Mechanism of stimulated Hawking radiation in a laboratory Bose-Einstein condensate. *Phys. Rev. A* **96**, 023616 (2017).
30. Tettamanti, M., Cacciatori, S. L., Parola, A. & Carusotto, I. Numerical study of a recent black-hole lasing experiment. *EPL* **114**, 60011 (2016).
31. Nozières, Ph. & Pines, D. *The Theory of Quantum Liquids* (Addison-Wesley, Reading, MA, 1990), Vol. II, Chap. 5.
32. Wang, Y.-H., Jacobson, T., Edwards, M. & Clark, C. W. Induced density correlations in a sonic black hole condensate. *SciPost Phys.* **3**, 022 (2017).
33. Corley S. & Jacobson, T. Black hole lasers. *Phys. Rev. D* **59**, 124011 (1999).
34. Volovik, G. E. Black hole and Hawking radiation by Type-II Weyl Fermions. *JETP Letters* **104**, 645-648 (2016).

**Acknowledgements**

We thank R. Parentani, N. Pavloff, A. Ori, G. Volovik, T. Jacobson, I. Carusotto and F. Sols for helpful discussions. This work was supported by the Israel Science Foundation.

**Author contributions**

J. R. M. d. N. and J. S. designed and built the experimental apparatus. J. R. M. d. N., K. G., and V. I. K. performed theoretical calculations. J. S. acquired the data. K. G., V. I. K., and J. S. analyzed the data. J. R. M. d. N. performed the numerical simulations. J. R. M. d. N. and J. S. prepared the manuscript with input from all authors.

**Competing interests**

The authors declare no competing financial or non-financial interests.

**Data Availability**

Source data are available for this paper. All other data that support the plots within this paper, and other findings of this study, are available from the corresponding author upon reasonable request.

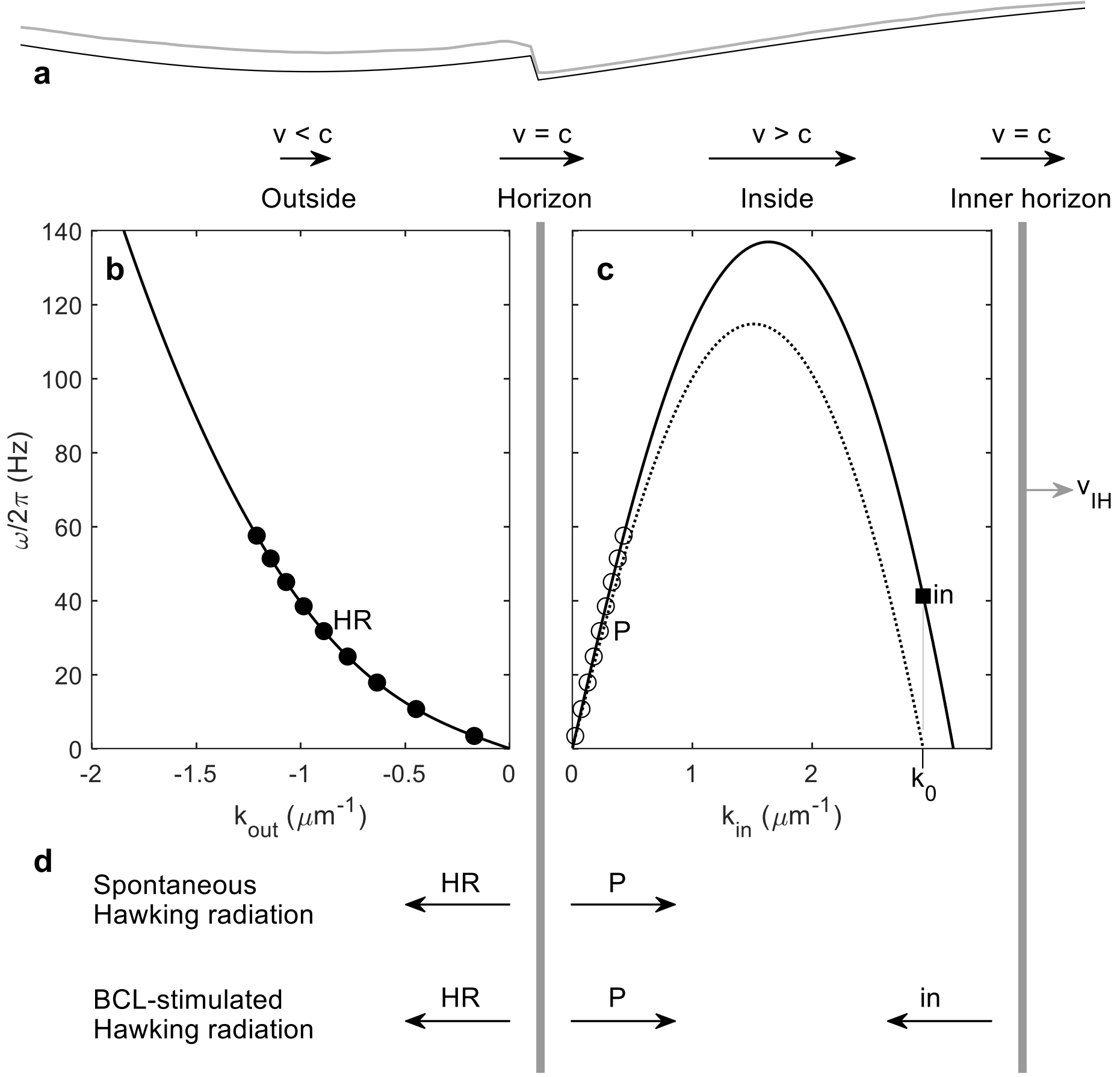

**Figure 1 | Spontaneous versus stimulated Hawking radiation. a,** The external potential including the step at the horizon. The gray curve is the flowing condensate. The curves for the spontaneous period are shown. **b,** The dispersion relation outside the analogue black hole. The values for a time of 312 ms are shown. Various outgoing Hawking modes HR are schematically indicated by circles. **c,** The dispersion relation inside the analogue black hole. The outgoing partner modes P are schematically indicated by circles. The square indicates the BCL mode incoming to the (outer) horizon. The inner horizon moves away from the outer horizon with velocity $v_{\mathrm{IH}}$. The dotted curve is the dispersion relation in the reference frame of the inner horizon. It has a zero-frequency mode at $k_0$. **d,** For spontaneous Hawking radiation, the population of incoming modes from either side of the horizon is zero. In BCL-stimulated Hawking radiation, incoming modes are produced at the inner horizon with a well-defined energy.

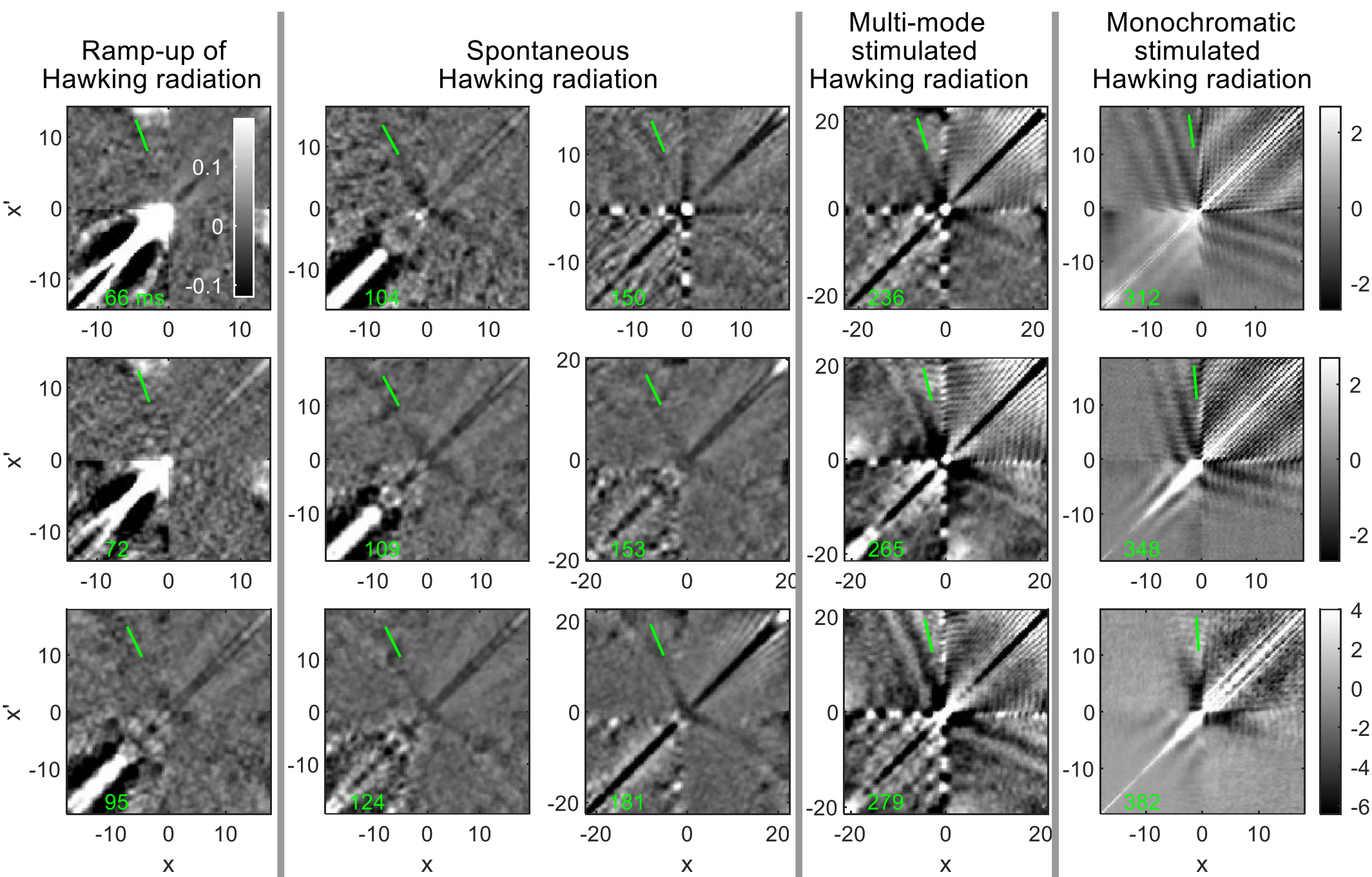

**Figure 2 | The observation of Hawking radiation at various times.** The correlation function $G^{(2)}(x, x')$ is shown. The time after the formation of the horizon is indicated. The grayscale for the first four columns is as in the first plot. The green line indicates the angle determined by the propagation speeds outside $(c - v)$ and inside $(v - c)$ the analogue black hole. The correlation functions in the first four columns have been filtered to reduce the noise. In the last column the signal is very large and no filtering has been applied, which improves the visibility of the features with high spatial frequency.

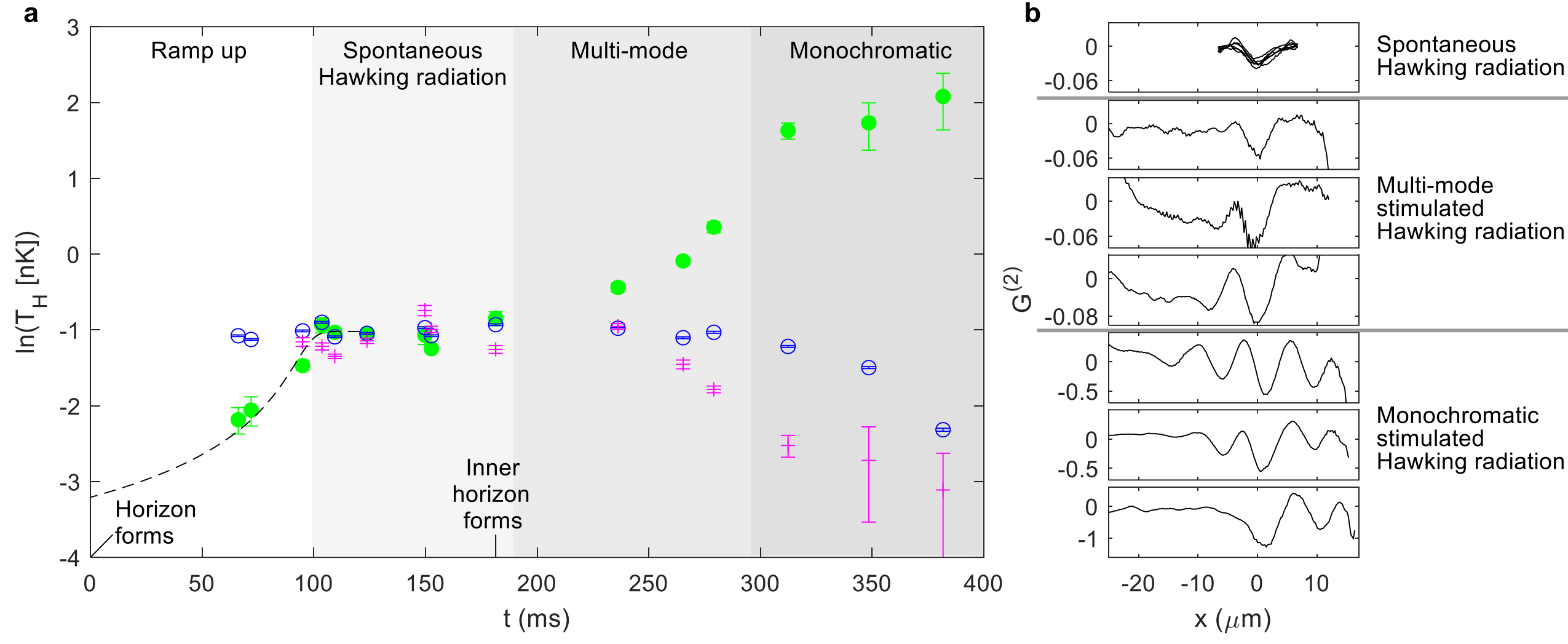

**Figure 3 | The evolution of the Hawking radiation. a,** The energy scales as a function of time, expressed as Hawking temperatures $T_{\mathrm{H}}$. The approximate formation times of the outer and inner horizons are indicated. The blue open circles indicate $T_{\mathrm{H}}$ predicted from the analogue surface gravity. The green filled circles indicate the quantity of emitted Hawking radiation. The magenta pluses indicate the width of the emitted spectrum of Hawking radiation. For a thermal spectrum at the predicted Hawking temperature, all 3 symbols agree. The dashed curve indicates the ramp up of the Hawking radiation in a real black hole during gravitational collapse. The error bars include the standard error of the mean and systematic errors. **b,** The profiles of the Hawking/partner correlation pattern seen in Fig. 2. The first panel shows the 6 profiles of the spontaneous period.

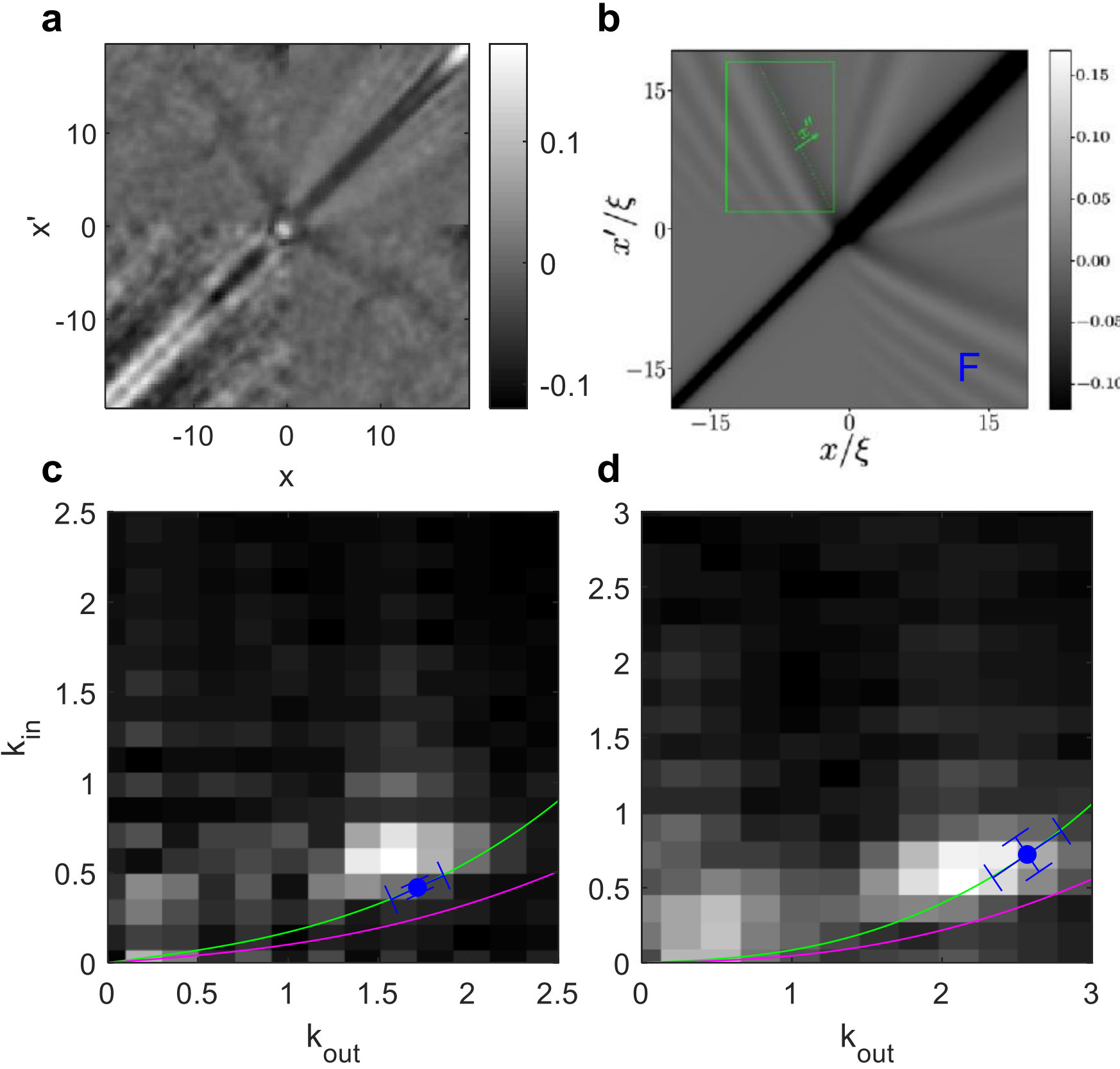

**Figure 4 | The spontaneous and monochromatic periods.** **a,** The average $G^{(2)}(x,x')$ for spontaneous Hawking radiation. The average of the 6 observations in the spontaneous period of Fig. 2 is shown. **b,** The theoretical prediction for an ideal step potential from Ref. 28. The dispersion fringes are indicated by F. $\xi$ is the healing length. In contrast to our work, Ref. 28 uses the green rectangle to compute the profile of $G^{(2)}(x,x')$. Reprinted figure with permission from M. Isoard and N. Pavloff, Phys. Rev. Lett. 124, 060401, 2020. Copyright (2020) by the American Physical Society. **c,** The 2-dimensional Fourier transform at 312 ms during the monochromatic period, computed within the upper left quadrant of the correlation function in Fig. 2. $k_{\mathrm{out(in)}}$ are the wavenumbers outside (inside) the analogue black hole. The green curve indicates the Hawking/partner pairs. The magenta curve indicates Hawking particles paired with positive energy particles directed into the analogue black hole in the free-falling frame. The green and magenta curves are derived from the measured dispersion relations (see Methods). The blue circle is the prediction for BCL stimulation. The error bars include the uncertainties in $v_{\mathrm{IH}}$, the flow velocities, and the speeds of sound. **d,** The Fourier transform at 348 ms.

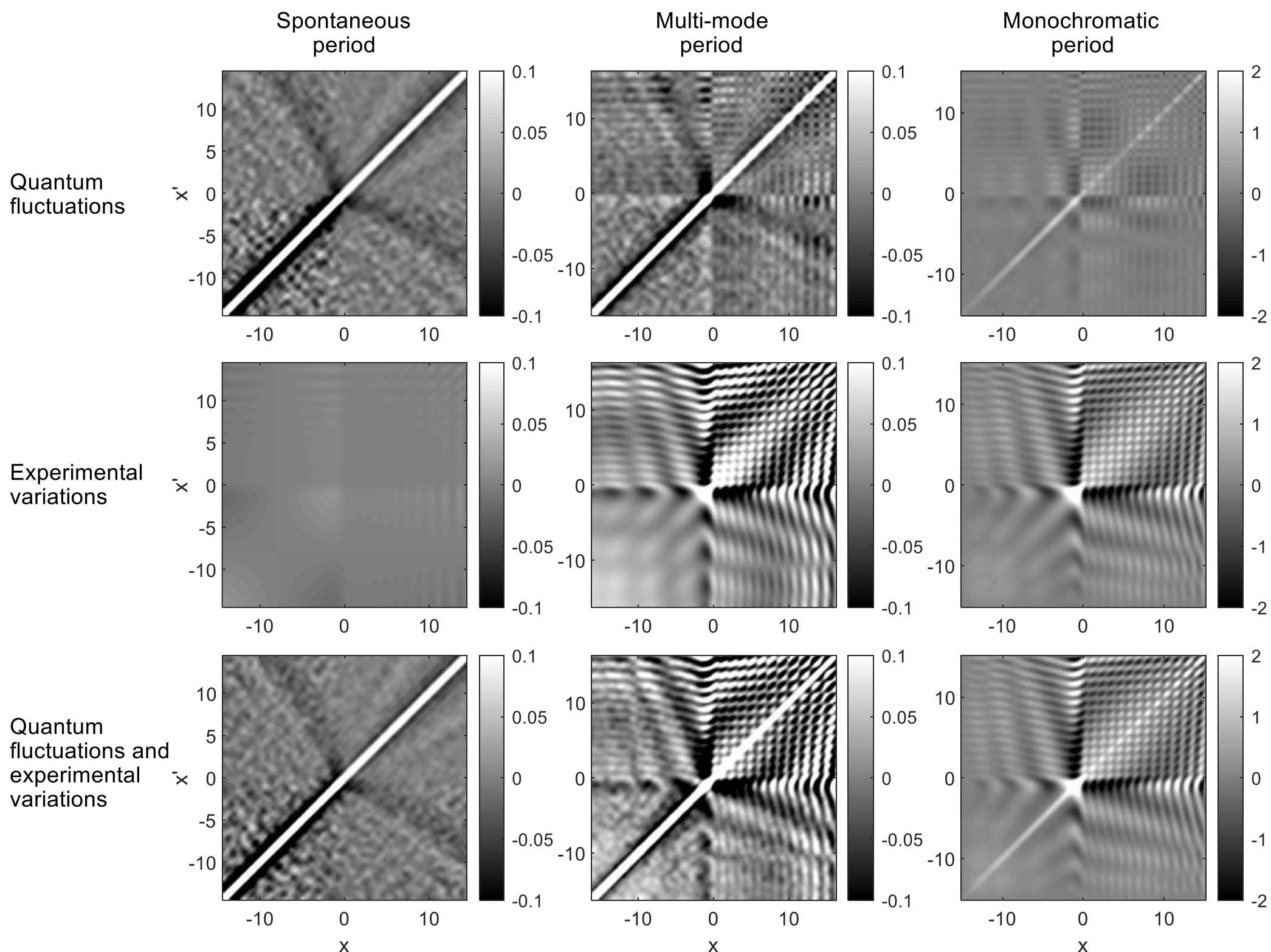

**Figure 5 | Numerical simulation of the experiment.** The correlation function $G^{(2)}(x, x')$ is shown. The first row includes quantum fluctuations only. The second row includes experimental variations between runs only. The third row includes both quantum fluctuations and experimental variations between runs. From left to right, the columns show the spontaneous, multi-mode, and monochromatic periods.

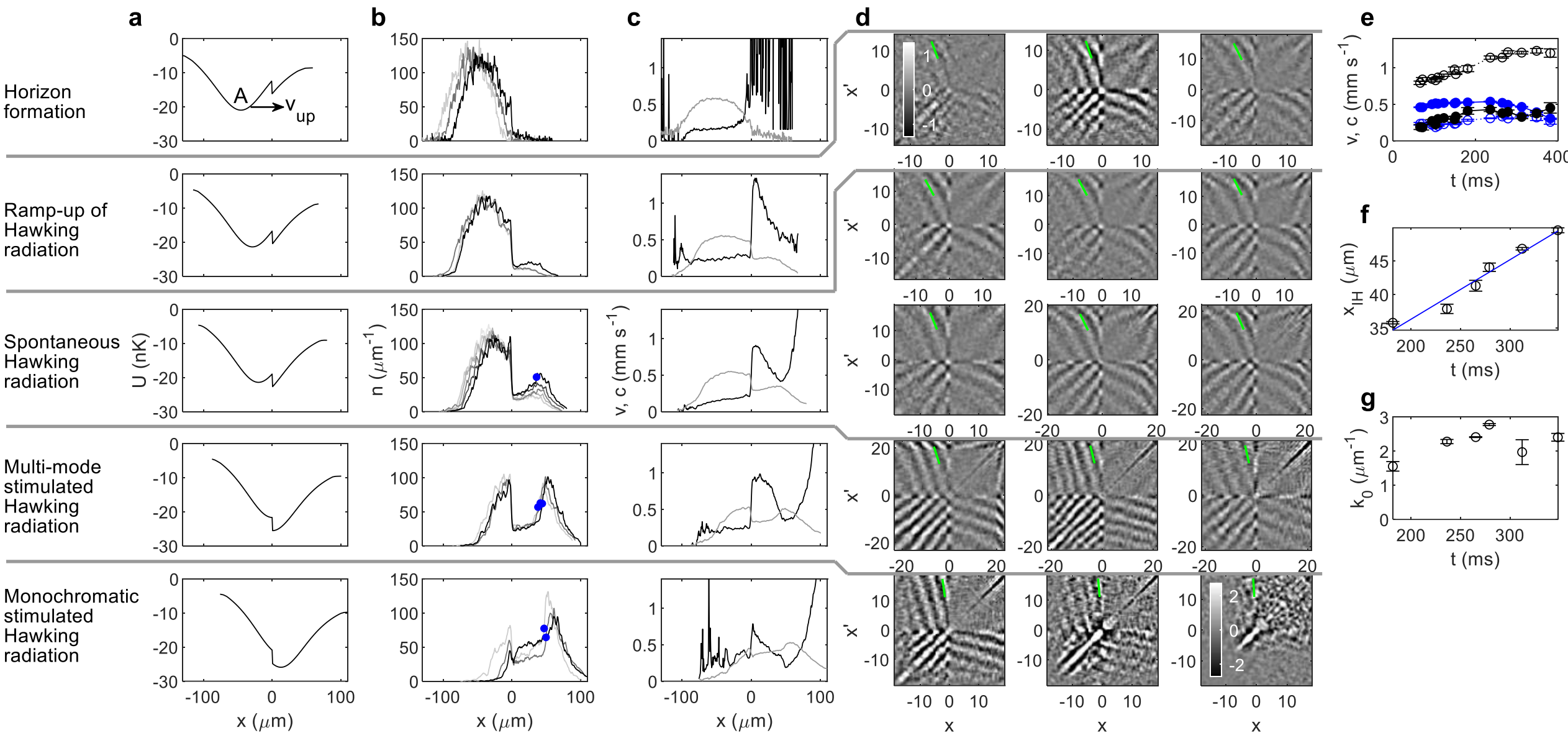

**Extended Data Figure 1 | The structure of the analogue black hole at various times.** The horizon frame is shown. The horizon is at $x = 0$. **a,** The average external potential $U(x)$. The Bose-Einstein condensate is initially in the minimum A. $v_{\mathrm{up}}$ is the applied velocity. **b,** Ensemble-average density profiles $n(x)$. Time increases from lighter gray to darker gray. The circles indicate the estimated position of the inner horizon. **c,** The profiles $v(x)$ (black curve) and $c(x)$ (gray curve) which determine the metric. **d,** Single modes from the preliminary oscillating horizon experiment. $G^{(2)}(x, x')$ for 30 Hz is shown. The times are the same as in Fig. 2. The grayscale is the same for all plots except the last. The green line indicates the angle determined by the propagation speeds outside $(c - v)$ and inside $(v - c)$ the analogue black hole. **e,** The time dependence of $v$ (open circles) and $c$ (filled circles) outside the analogue black hole (blue) and inside (black), obtained from the measured dispersion relations. The error bars indicate the standard error of the mean. **f,** The position $x_{\mathrm{IH}}$ of the inner horizon from **b**. The error bars indicate the standard error of the mean. The linear fit gives $v_{\mathrm{IH}} = 0.09(1)$ mm $s^{-1}$. **g,** The wavenumber $k_0$ of the short-wavelength superluminal wave between the horizons, found from the location of the peak in the static structure factor computed from the density profiles inside the analogue black hole. The error bars indicate the standard error of the mean.

## Methods

The atomic Bose-Einstein condensate is initially in the potential minimum at point A in the first panel of Extended Data Fig. 1a. In the laboratory frame, the step in the potential is moved to the left with a constant applied velocity $-v_{\mathrm{up}}$. In the horizon reference frame in which the step potential is stationary, the potential minimum moves to the right with a constant velocity of $v_{\mathrm{up}}$ = 0.16 mm $s^{-1}$. The condensate pours over the step at $x = 0$ and is accelerated to supersonic speeds. The resulting ensemble-average density profiles are shown in Extended Data Fig. 1b. In the horizon formation period, each curve is an average of 3 repetitions of the experiment. For the other periods, each curve is an average over an ensemble of between 4000 and 11,000 repetitions of the experiment. Relative to the formation time of the horizon, the times shown are -83, -17, and 50 ms (horizon formation period), 66, 72, and 95 ms (ramp-up period), 104, 109, 124, 150, 153, and 181 ms (spontaneous period), 236, 265, and 279 ms (multi-mode period), and 312, 348, and 382 ms (monochromatic period).

We estimate the flow velocity for various pairs of density profiles $n_1$ and $n_2$ for times $t_1$ and $t_2$ within each panel of Extended Data Fig. 1b by the one-dimensional continuity equation $v(x) = -[2/(n_1+n_2)] \int dx\,(n_2-n_1)/(t_2-t_1)$ (Ref. 35). The speed of sound is computed for each profile by the relation[25]

$$c(x) = \sqrt{\frac{2\hbar a \omega_r n}{m}} \sqrt{\frac{1+3na/2}{(1+2na)^{3/2}} - \frac{\hbar \omega_{r_0}}{2U_0}}$$

By averaging various $v(x)$ and $c(x)$ within each panel, we obtain the curves shown in Extended Data Fig. 1c. The horizon is the point where the curves intersect near $x = 0$. In the last two panels, the curves cross again at larger $x$, forming the inner horizon. Due to various sources of error, $v(x)$ in Extended Data Fig. 1c is approximate and qualitative only. More accurate values of spatial averages of $v$ and $c$ are obtained by measuring the dispersion relation via our oscillating horizon technique[24,25]. In this preliminary experiment, the step potential at the horizon is caused to oscillate with a definite frequency which generates waves outside and inside the analogue black hole. The step oscillates with an amplitude of 0.5 μm, with the exceptions of the second and the latest times which have amplitudes of 1 μm and 1.5 μm, respectively. For the 4 latest times, the oscillation of the step potential begins at 233 ms, 266 ms, 299 ms, and 366 ms. For all other times, the oscillation begins at -166 ms. A random phase is chosen for each of the many runs, and the density-density correlation function is computed, as shown for 30 Hz in Extended Data Fig. 1d. The sinusoidal single-mode pattern gives the wave vectors outside and inside the analogue black hole corresponding to the applied frequency. Fitting Doppler-shifted Bogoliubov spectra to the resulting dispersion relations gives the values of $v$ and $c$ shown in Extended Data Fig. 1e.

---

35. Lahav, O., Itah, A., Blumkin, A., Gordon, C., Rinott, S., Zayats, A. & Steinhauer, J. Realization of a Sonic Black Hole Analog in a Bose-Einstein Condensate. *Phys. Rev. Lett.* **105**, 240401 (2010).